# An Energy Efficient Decoding Scheme for Wireless Body Area Sensor Networks

O. Rehman, N. Javaid, A. Haider, N. Amjad, A. A. Awan, M. Qamar, Z. A. Khan*, U. Qasim[#]

COMSATS Institute of Information Technology, Islamabad, Pakistan.
*Internetworking Program, Faculty of Engineering, Dalhousie University, Halifax, Canada.
[#]University of Alberta, Alberta, Canada.

**ABSTRACT**

One of the major challenges in Wireless Body Area Networks (WBANs) is to prolong the lifetime of network. Traditional research work focuses on minimizing transmit power; however, in the case of short range communication the consumption power in decoding is significantly larger than transmit power. This paper investigates the minimization of total power consumption by reducing the decoding power consumption. For achieving a desired Bit Error Rate (BER), we introduce some fundamental results on the basis of iterative message-passing algorithms for Low Density Parity Check Code (LDPC). To reduce energy dissipation in decoder, LDPC based coded communications between sensors are considered. Moreover, we evaluate the performance of LDPC at different code rates and introduce Adaptive Iterative Decoding (AID) by exploiting threshold on the number of iterations for a certain BER ($10^{-4}$). In iterative LDPC decoding, the total energy consumption of network is reduced by $20-25\%$.

## I. INTRODUCTION

A WBAN is a special case of Wireless Sensor Networks (WSNs) that enables remote monitoring of various sensors in several environments. One of the most studied application of WBANs is health care monitoring, where few or large number of patients can be observed, diagnosed, and prescribed remotely. Smart sensor nodes can be connected to various parts of the body or fabricated inside the clothes to transmit the information to a base station. Thus WBANs emerged as a promising alternative for traditional wired network systems. Recent advances in wireless communications, signal processing and digital electronics enabled the development of tiny wireless sensors with small batteries and the capability to sense, process and communicate with each other. In WBANs, area occupation and energy consumption are important aspects.

The issue of energy consumption is very critical and needs to be minimized at the design levels like modulation/demodulation, Media Access Control (MAC) protocols, routing protocols and error correction coding. Reliable communication primarily depends on the applications and user specified constraints such that, receiver should embed the error correction strategies to recover original data. In a particular error correction code, the amount of energy spent to perform channel coding should be significantly less than the energy saved by transmitter.

Fig. 1 shows a Wireless Body Area Network(WBAN). To tackle erroneous transmission in WBAN, Automatic Repeat Request (ARQ) strategy is usually implemented by using ARQ protocol. In this protocol sensor nodes implement error detection code for identifying corrupted frames and then requests for a retransmission. However, this technique is inefficient when we deal with energy and delay. For a lossy channel Error Correcting Codes (ECC) are commonly used to reduce the number of retransmissions. Incase of packet encoding and decoding, ECC requires more processing power at sensor nodes. To recover errors before a transmission request is sent HARQ protocol is used.

For ECC error correction is directly related with complexity. Powerful codes increases the processing energy consumption at the receiver. So, there is trade-off between transmission and processing energy consumption at decoder. FEC use simple error correction techniques like Reed-solomon, Bose and Ray Chaudhuri (BCH) and conventional codes and their ability to correct errors is obtained by introducing high redundancy in transmitted data. Coding and decoding process introduces delay in delivering packets to the base station. For bad quality of communication channels, LDPC codes are used to achieve low BER with low redundancy.

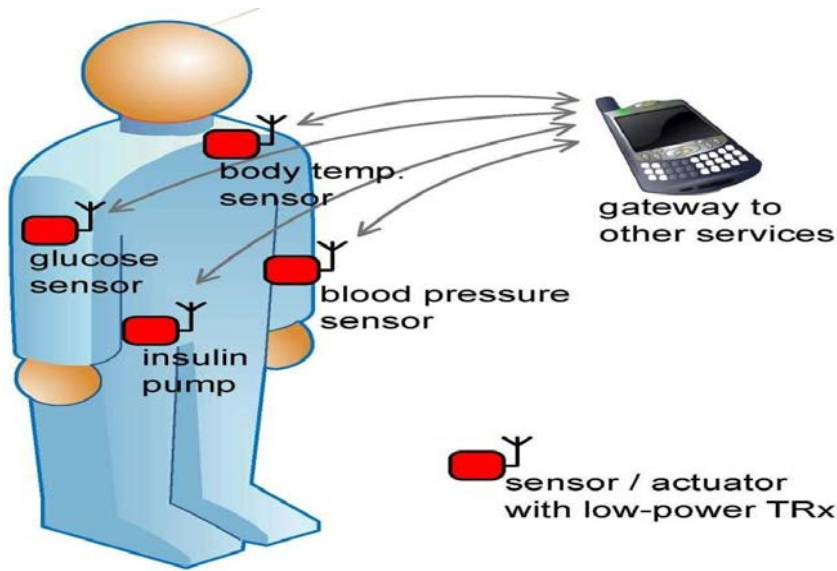

Fig. 1: Wireless Body Area Sensor Network

ECC introduces redundancy into an information sequence $U$ of length $k$ by adding extra parity bits, based on various combination of bits of $u$, to form a codeword $x$ of length $n_c > k$. The redundancy provided by these extra $n_c - k$ parity bits allows the decoder to possibly decode noisy received bits of $x$ correctly which if uncoded, would be demodulated incorrectly. Compared to an uncoded system, ECC over noisy channel provides better BER performance for same Signal to Noise Ratio (SNR). In comparison to an uncoded system to achieve a certain BER at required SNR for coding and decoding algorithm is known as the coding gain for that code and decoding algorithm. Typically there is tradeoff between coding gain and decoder complexity. Very long codes provide higher gain. However, require larger decoders with high power consumption, and similarly for more complex decoding algorithms. ECC are mainly categorized into two types:

(1) Block Codes have fixed length $nc$ that include $n_c - k$ parity bits. These codes are decoded one codeword at a time. (2) Conventional codes have a code rate $\frac{k}{n_c}$, where $k$ are number of input bits and $n_c$ are number of output bits that are decoded in a continuous stream of length $L \gg n_c$. Block codes include Hamming codes, repetition codes, Reed-Solomon (RS) codes. The terminology ($n_c$, $k$, $d_{min}$) represents a code of length $n_c$ with input bits of length $k$ and minimum distance $d_{min}$.

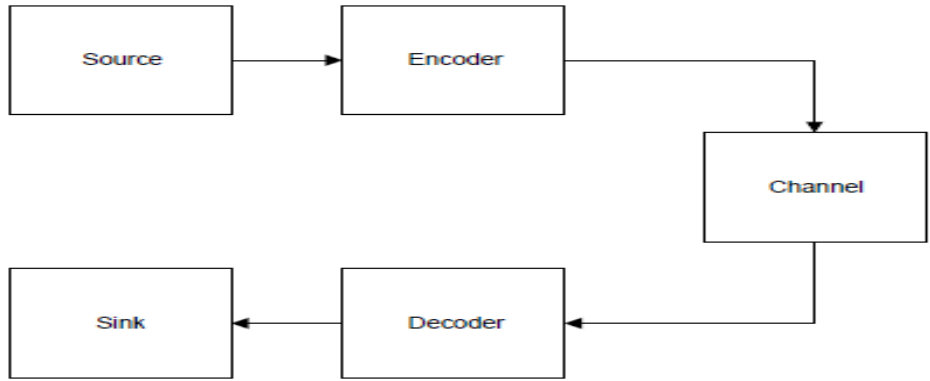

Fig. 2: Basic Communication Model

ECC are also categorized in terms of decoding algorithms:
**1.** Non iterative decoders are divided into two types. (a) Hard decision decoder is a threshold detector stage of a decoder where decision is based on binary signaling that includes 1 or 0. (b) In soft decision decoders more than one voltage levels are considered for a single bit. These Soft information are calculated in the form of Log Likelihood ratios (LLRs).
**2.** For each iteration, iterative decoders require soft information as an input. These algorithms provide higher coding gain. However, they require complex decoder and consume significant power. LDPC codes are applicable in WSNs because of Joint source coding and channel coding. Basic communcation model has been depicted in Fig. 2.

The rest of the paper is organized as follows: In section 2, we briefly discuss related work. Section 3 focuses on our motivation for research work. In Section 4, we discuss LDPC Codes. Section 5 presents AID scheme for LDPC. Simulation results are shown in Section 6 and Section 7 concludes our research work.

## II. BACKGROUND AND MOTIVATION

This work is an extension of [2]. Recent research in WSNs mostly deals with the aim to maximize energy efficiency ([3]). In this regard, some authors worked at routing layer ( [4], [5], [6]) while others ( [7], [8], [9]) explored MAC layer. However, our directions are more focused on different decoding schemes to achieve maximum energy efficiency. In [10], P. Grover *et al.* investigate the perspective of interference on decoding power. They suggest that in short range communication the transmit power is smaller than decoding power and uncoded transmission requires more transmit power than coded transmission. Andrea *et al.* designed LDPC decoder architecture for WSN. Different LDPC codes are considered to analyze the energy saving w.r.t un-coded communication, depending on distance, BER and Environment [11]. S.L.Howard *et al.* calculate critical distance *dc* at which the decoder's energy consumption per bit is equal to transmit energy per bit. In comparison to an un-coded system, authors provide results for *dc* in different environments over a wide frequency range [12]. Marcelo *et al.* investigate the tradeoff between transmission and processing energy consumption in sensor nodes by employing convolutional codes. For each sensor node, authors select appropriate complexity for ECC to prolong network lifetime [13]. Z. Hajjarian *et al.* define the relationship between the number of quantization bits and decoder's energy consumption using LDPC in WSN. Decoder's complexity is reduced by replacement of functional blocks with look up tables. They suggest that LDPC codes are more energy efficient than conventional and block codes. Using iterative decoding, the network lifetime is increased up to four times with regular LDPC codes [14]. In [15], authors propose a packet error reduction technique for reliable communication in WBAN. Firstly, they calculate Received Signal Strength (RSS) around the human body. Secondly, based on these calculations of RSS, they calculate Packet Error Rate (PER). LT codes are considered for

WBAN to reduce the PER. In [16], authors use rateless codes to provide an adaptive duty cycling for power management. Analytical results show that with same structure used for WSNs, upto 80 percent of energy is saved as compared to IEEE 802.15.4 physical layer standard. Z. H. Cai *et al.* propose an efficient early stopping method to reduce number of iterations for LDPC decoders. This method is very efficient at low SNR [17]. Energy efficiency plays a vital role in enhancing the lifetime of a network. As in [18] – [40], many efforts have been made to improve the energy efficiency and, hence, the improving the lifetime of network.

WBAN consists of small sensors with limited battery power. In this paper, our objective is to increase the network lifetime or to minimize the energy consumption of network. Reliability is another primary requirement for communication. The level of reliable communication depends on applications and user specific constraints. Error correction schemes provide reliable communication between transmitter and receiver by reducing $P_b$. In practical situations, such as WBAN there are two ways to reduce $P_b$.

**1.** Increasing the transmit power improves the SNR:
For Additive White Gaussian Noise (AWGN) channel, $p(b)$ for BPSK modulation is expressed as:

$$p_b = Q\sqrt{\frac{2E_b}{N_0}} \qquad (1)$$

where $Q$ is a scaled form of the complementary Gaussian error function.

**2.** Use complex decoder by increasing the decoding power:
In the case of WBAN an increase in transmit power is not sufficient because increase in transmit power damages the human tissues. Another problem is the existence of small distance between receiver and transmitter.

The transmit power increases the Signal to Interference and Noise Ratio (SINR). In this case how we can improve SINR? This is done by increasing the distance between transmitters. However, in WBAN it is not possible because of limited area. In any communication system a tradeoff between transmit power and decoding power exists. WBAN requires low transmitting power so, there is need to implement energy efficient decoder to reduce the decoding power consumption to enhance the network lifetime. We use LDPC, because its performance for low BER communication is same as complex decoders such as viterbi decoding algorithms.

### III. LDPC CODES

LDPC codes are a class of linear block codes, their parity check matrix contains only a few number of 1s in comparison of 0s. The main advantage of this code provides decoding performance very close to the capacity and linear time complex algorithms for different AWGN and fading channels. Like all linear block codes, LDPC codes can be represented in two different possibilities. They can be described via graphical representation and matrices. LDPC codes are represented as as Tanner Graph that contain two set of nodes: Check Nodes (CNs) and Variable Nodes (VNs). VNs are associated with N bits of codeword and CNs corresponds to M parity-check constraints. Edges in Tanner graph correspond to $1's$ in H and exchange of information along these edges as shown in Fig. 3.

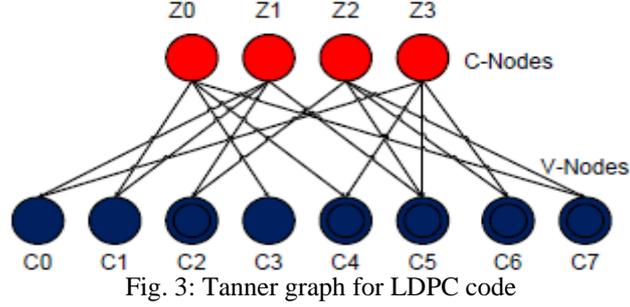
Fig. 3: Tanner graph for LDPC code

Belief Propagation (BP) algorithm is commonly used to decode LDPC codes. VNs receive LLR from the channel and update it according to parity check matrix computed at CNs. This process is iterated until iterations reach a maximum level. This criteria helps in successful decoding of a codeword. BP algorithm has two main scheduling schemes: two phase scheduling and layered scheduling. In layered scheduling, parity check constraints are grouped in the form of layers. Each layer is associated with a component code. One iteration is complete when all layers are successfully decoded. To achieve a reliable pattern of information, the whole process is iteratively repeated up to several times.

Let $S_j$ represents the LLR of the bit in column $j$ of $H$. LLR is initialized to the corresponding received soft value. For each parity constraint $m$ in a given layer, the following operations are executed [5]:

$$H = \begin{vmatrix} 0 & 1 & 0 & 1 & 1 & 0 & 0 & 1 \\ 1 & 1 & 1 & 0 & 0 & 1 & 0 & 0 \\ 0 & 0 & 1 & 0 & 0 & 1 & 1 & 1 \\ 1 & 0 & 0 & 1 & 1 & 0 & 1 & 0 \end{vmatrix}$$

$$Q_{mj} = S_j^{(old)} - R_{mj}^{(old)} \qquad (2)$$

$$A_{mj} = \Sigma_{\eta \varepsilon N_m, n \neq j} \, \psi(Q_{mn}) \qquad (3)$$

$$s_{mj} = \Sigma_{\eta \varepsilon N_m, n \neq j} \, sgn(Q_{mn}) \qquad (4)$$

$$R_{mj}^{(new)} = -s_{mj} \cdot \psi(A_{mj}) \qquad (5)$$

$$S_j^{(new)} = Q_{mj} + R_{mj}^{(new)} \qquad (6)$$

$S_j^{(old)}$ is the extrinsic information received from previous layer and updated in equation (5). This information is propagated to succeeding layer. $R_{mj}^{(old)}$ is used to compute equation (1) then updated in equation (4). Term, $R_{mj}^{(new)}$ is used again in following iteration. $N_m$ in equations (2) and (3) represents set of all bits connected to parity constraint m. Whole operation of LDPC decoder is given in Fig. 4.

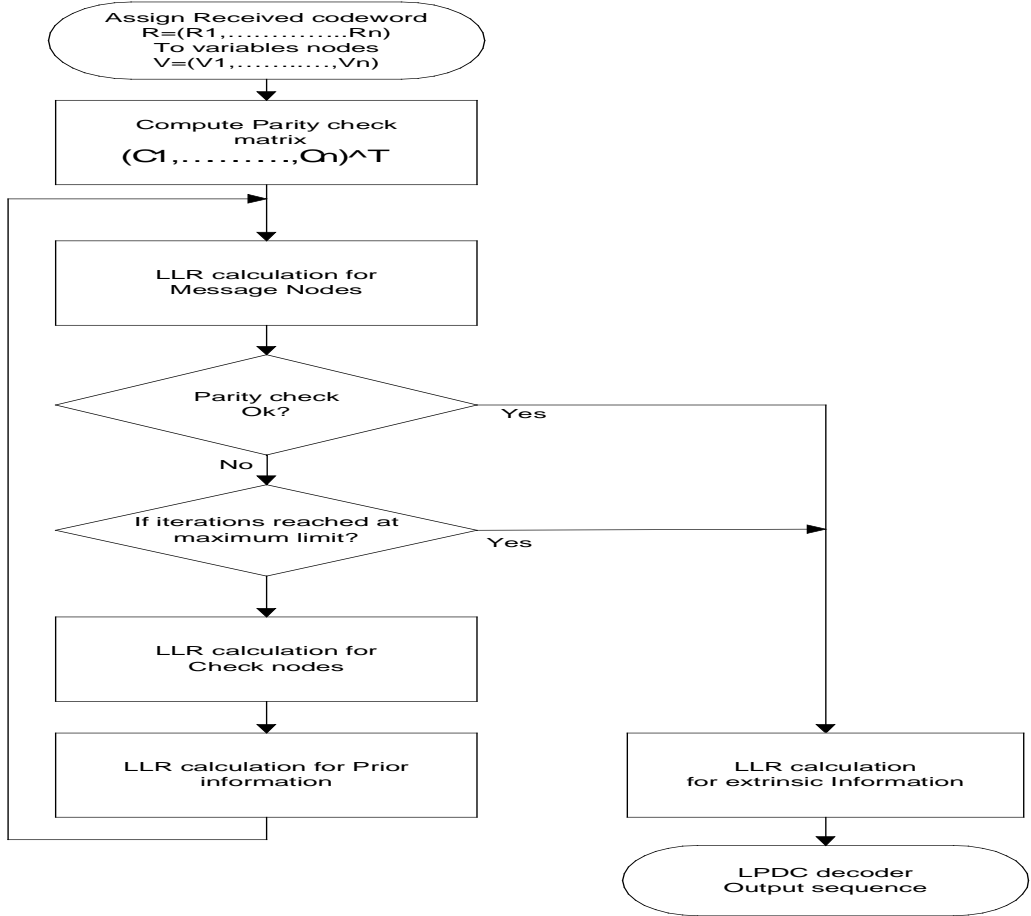

Fig. 4: Flow chart for LDPC decoder

## IV. SENSOR NODES ENERGY MODEL

For encoded data of transmission, the energy consumption per bit of a node is the sum of radio energy per bit $E_r$ and computation energy per bit $E_{comp}$. Radio energy is sum of SLEEP, TRANSIENT, and ON mode energies. The transceiver is off all the times and keeping it ON only when data is transmitting or receiving.

The radio energy per bit for transmitting *N* bits is calculated as in [12]:

$$E_r = \frac{p_{sleep}T_{on} + p_{tran}T_{tr} + p_{on}T_{on}}{N} \qquad (7)$$

Power consumed in radio during ON-mode is the sum of transmitting power $p_t$ and power consumption in circuitry $p_{ckt}$. The radio energy consumption is:

$$E_r = \frac{(p_t + p_{ckt})T_{on}}{N} \qquad (8)$$

Transmission power in free space is expressed by using Friis equation:

$$p_t = \frac{p_r}{G_r G_t} \times \left(\frac{4\pi}{\lambda}\right)^2 d^n \qquad (9)$$

The received power is calculated as:

$$p_r = SNR_{uncod} bB \frac{N_0}{2} NF \qquad (10)$$

Where, $SNR_{uncod}$ is SNR for transmitting uncoded data, b is the number of bits per modulation symbol, B is bandwidth, $\frac{N_0}{2}$ is noise spectral density and NF is Noise Figure.
For coded data received power is:

$$p_r = SNR_{coded} bB \frac{N_0}{2} NF \qquad (11)$$

Transmission energy for coded data frame is expressed as:

$$E_{trans}^{frame} = E_{dec}^{bit} . f_{size}^C \qquad (12)$$

The required processing energy to decode a bit is $E_{dec}^{bit}$. Therefor the energy required to decode a frame $E_{dec}^{frame}$ is computed as:

$$E_{dec}^{frame} = E_{dec}^{bit} . f_{size}^C . r (J/F) \qquad (13)$$

where *r* is the coding rate.
There is tradeoff between transmitting power and decoding power. To reduce the total power consumption, transmitting power must strictly be larger than Shannon limit. In [9] the Shannon capacity limit for both coded and uncoded systems is shown in Fig. 5.

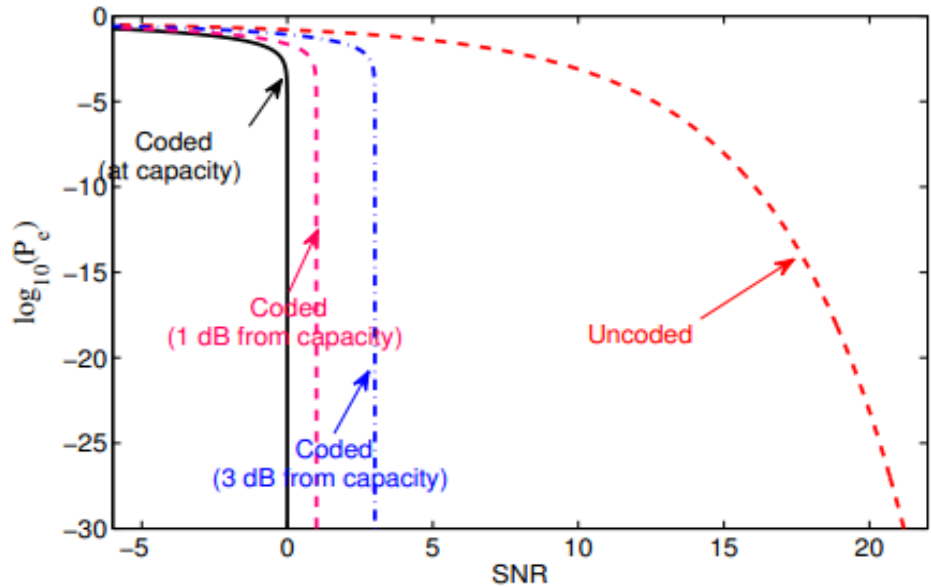

Fig, 5: Shannon Capacity estimation for coded and uncoded system

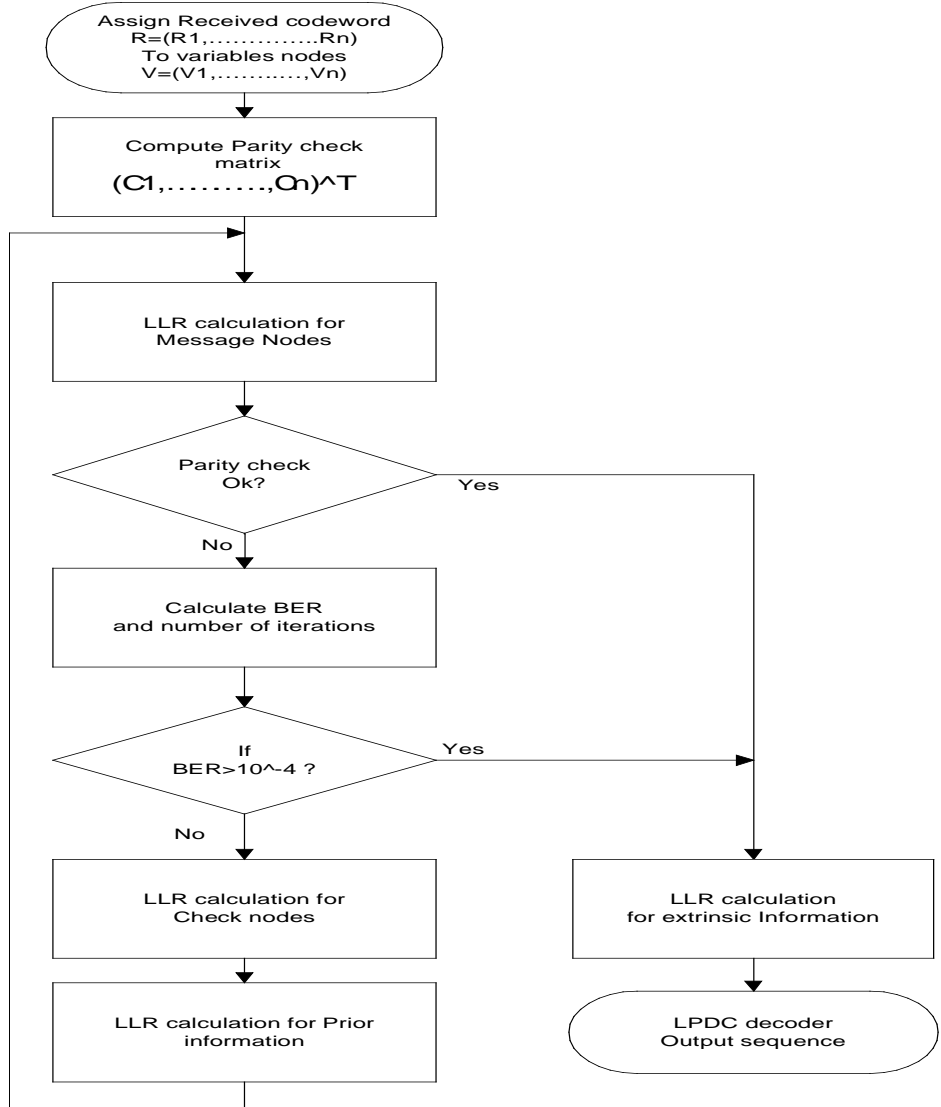

Fig. 6: Flow chart of AID for LDPC

## V. AID

In this paper, we propose an efficient early stopping method to reduce the number of iterations for LDPC decoders. This method is efficient at low SNR.

AID predicts the threshold at number of iterations for undecodable blocks. Another advantage is reduction in processing delay. LDPC codes yield excellent performance when they are decoded with iterative message passing algorithms. The performance of an iteratively decoded LDPC code is typically reported in terms of $p_e$ for a specific SNR value. For high SNR values, the value $p_e$ is small that requires a large number of iterations in order to obtain reliable estimate. In AID, we introduce a threshold on the number of iterations after achieving a certain BER ($10^{-4}$). The whole process of AID is shown in Fig. 6.

We assume that each PE consumes a fixed $E_{node}$ joules of energy per iteration. We assume that the $R_{dec}$ is equal to data rate $R_{data}$ measured in information bits per second. The power received at distance *x* meter is given in Eq [9]. In Eq [13] the $E_{dec}$ is the energy consumed in decoder operation. For any rate $R_{data}$ the average probability of error $(p_e) \to 0$. Let the number of decoding iterations be denoted by *l*, the number of computational nodes can be lower bounded by m, the number of received channel outputs. Since each node consumes $E_{node}$ joules of energy in each iteration, the decoding energy $E_{dec}$ is lower bounded in [2] as: $E_{dec} \geq E_{node} \times m \times l$. We assume channel encoding is free. This result is the following lower bounds on the weighted total power is $E_{tot} \geq p_T + \frac{E_{node} \times m \times l}{T_{dec}}$, and $E_{tot} = P_L P_R + \frac{E_{node} \times m \times l}{T_{dec}}$. Where $T_{dec} = \frac{k}{R_{dec}}$ is the time consumed in decoding. Thus, $p_{total} \geq P_L P_R + \frac{E_{node} \times m \times l \times R_{dec}}{k}$ and $p_{total} = P_L P_R + \frac{E_{node} \times l \times R_{dec}}{R_c}$, where, $P_L$ is power loss and $P_R$ is received power.

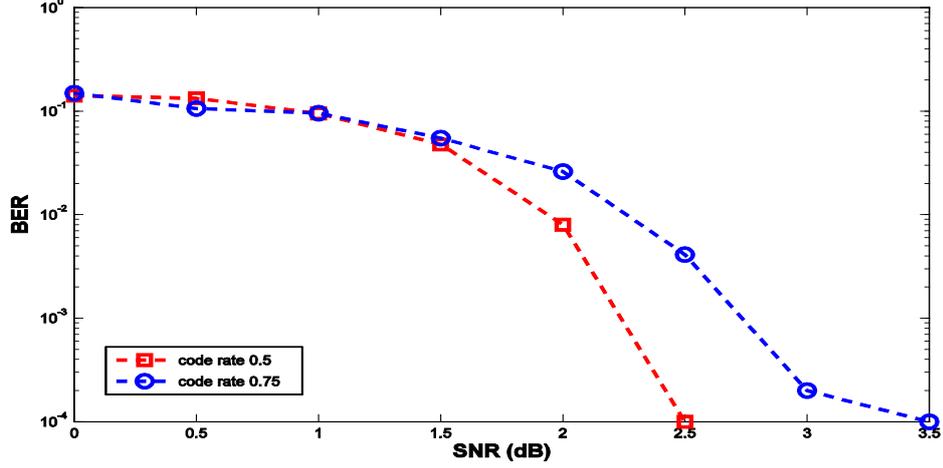

Fig. 7: Relationship between SNR and BER using LDPC

## VI. RESULTS AND DISCUSSIONS

LDPC code with Code Rate ($R_c$) has been used in simulation over AWGN channel with an iterative probabilistic decoding algorithm. The bandwidth is large (3GHZ) and the throughput is 1.5 Gbps. The power of a PE is 10 pico-Joules. For indoor environment, path-loss exponent are assumed to be 3 and noise figure is 3dB. The transmitter power is in few milli Watts for 3m distance between sensor and base station. If the transmit power $P_T$ is extremely close to that required for channel capacity then large number of iterations $l$ are required. Large number of iterations consume high decoding power. Encoder should have larger power as compared to its Shannon limit in order to minimize decoding power consumption.

Simulations parameters are shown in Fig. 10.

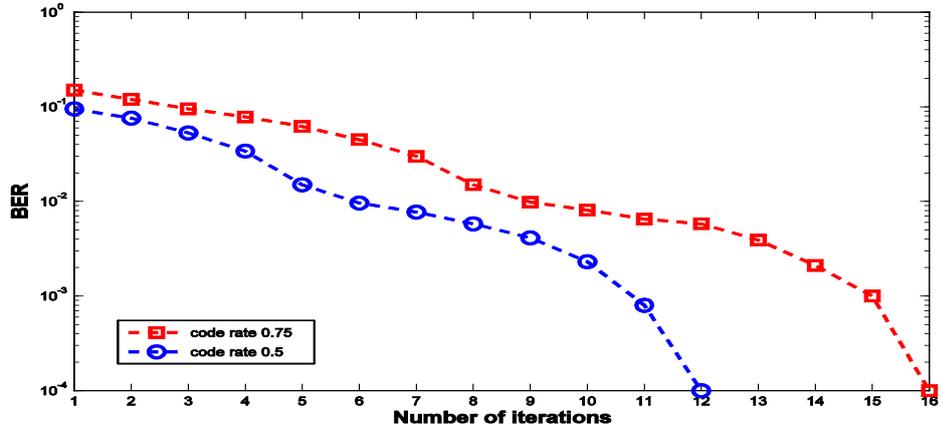

Fig. 8: Iterations in AID for desired BER($10^{-4}$)

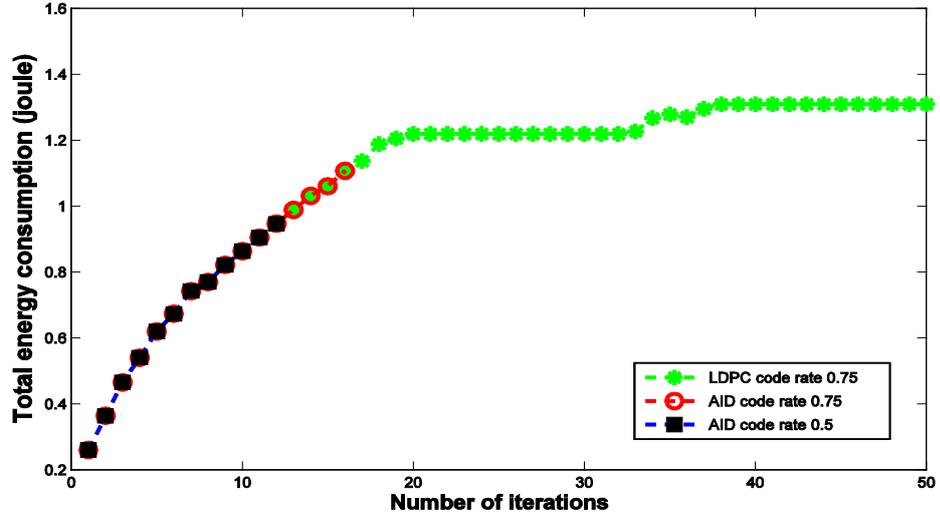

Fig. 9: Energy Consumption in AID

| PTr | 144mW |
|---|---|
| w | 1 |
| pathloss Exonent | 1 |
| d | 3m |
| Boltzmann Constant | $1.38065 * 10^{-23}$ |
| temperature | 300 |
| Rdata | 1.5Gbps |
| Rdec | 1.5Gbps |
| Enode | $10^{-12}$ |
| degree | 2 |
| Rc | 0.5 |
| LDPC | 7,4 |
| frequency | 60 GHZ |

Fig. 10: Simulation Parameters

Fig. 7 , shows the relationship between the SNR and BER, as we increase the SNR, BER decreases. By imposing five iterations ($l = 5$), it is observed that targeting a specific BER of $10^{-4}$ the requried SNR is 3.5 at $R_c$ 0.75 and SNR 2.5 at $R_c$ 0.5. Fig. 8 , shows the relationship between number of iterations and BER. For a fixed SNR (1.5dB) our target is to achieve a BER of $10^{-4}$ and calculate the number of iterations till desired BER of $10^{-4}$ is achieved.`

Results show that at $R_c$, 0.75 the required iterations are 16 and at $R_c$ 0.5, 12 iterations are required.

Fig. 9 , shows the relationship among the total number of iterations and total power consumption by using LDPC decoder and AID. For the case of LDPC code, we use 50 iterations at a code rate of 0.75. There is no threshold for achieving a certain value of $Pe$. As $Pe \rightarrow 0$ the total power consumption $p_T \rightarrow \infty$. Simulation results show that $p_T$ increases by increasing the number of iterations. For 50 iterations $p_T$ is 1.3 joule. In

AID, our requirement is to achieve a BER of $10^{-4}$, it is achieved at 16th iteration. Hence, we save the energy consumption of 35 iterations and also minimize the processing delay of decoder. This scheme is equally performed well for such applications in which we cannot afford delay. Simulation results show that the total energy consumption of error correction code is exactly 1.05 joule. With the code rate of 0.5, energy consumed is 0.5 joule in total. We save 22% of total energy as compared to LDPC codes. Hence, AID defines threshold on the number of iterations for a desired BER. Benefit of this decoding scheme is that it reduces the processing time of decoder and decoding energy consumption in WBAN.

## VII. CONCLUSION

In this paper, we proposed AID Scheme to reduce the decoding power consumption and to prolong the lifetime of WBANs. To achieve BER to ($10^{-4}$), we calculate the total number of iterations required for LDPC decoder. As $p_b$ approaches 0, iterations approach infinity and iterations stop after achieving a desired BER. We reduce decoding energy consumption by stoping the iterations after achieving a certain BER. By using this scheme, $20 - 25\%$ of total energy consumption is reduced.

**REFERENCES**

[1] M. Aslam, N. Javaid, A. Rahim, U. Nazir, A. Bibi, Z. A. Khan, Survey of Extended LEACH-Based Clustering Routing Protocols for Wireless Sensor Networks, 5th International Symposium on Advances of High Performance Computing and Networking (AHPCN-2012)in conjunction with 14th IEEE International Conference on High Performance Computing and Communications (HPCC-2012), 25-27 June, Liverpool, UK, 2012.

[2] N. Javaid, O. Rehman, N. Alrajeh, Z. A. Khan, B. Manzoor, S. Ahmed, "AID: An Energy Efficient Decoding Scheme for LDPC Codes in Wireless Body Area Sensor Networks", 4th International Conference on Emerging Ubiquitous Systems and Pervasive Networks (EUSPN-2013), Niagara Falls, Ontario, Canada.

[3] N. Javaid, A. Bibi, K. Dridi, Z. A. Khan, S. H. Bouk, "Modeling and Evaluating Enhancements in Expanding Ring Search Algorithm for Wireless Reactive Protocols", 25th IEEE Canadian Conference on Electrical and Computer Engineering (CCECE2012), Montreal, Canada, 2012.

[4] K. Latif, A. Ahmad, N. Javaid, Z. A. Khan and N. Alrajeh, "Divide-and-Rule Scheme for Energy Efficient Routing in Wireless Sensor Networks", The 4th International Conference on Ambient Systems, Networks and Technologies (ANT 2013), 2013,
Halifax, Nova Scotia, Canada, Procedia Computer Science, Volume 19, 2013, Pages 340-347.

[5] A. Ahmad, K. Latif, N. Javaid, Z. A. Khan and U. Qasim, "DENSITY CONTROLLED DIVIDE-AND-RULE SCHEME FOR ENERGY EFFICIENT ROUTING IN WIRELESS SENSOR NETWORKS", 26th IEEE Canadian Conference on Electrical and Computer Engineering (CCECE2013), Regina, Saskatchewan, Canada, 2013.

[6] M. B. Rasheed, N. Javaid, Z. A. Khan, U. Qasim and M. Ishfaq, "E-HORM: AN ENERGY-EFFICIENT HOLE REMOVING MECHANISM IN WIRELESS SENSOR NETWORKS", 26th IEEE Canadian Conference on Electrical and Computer Engineering (CCECE2013), Regina, Saskatchewan, Canada, 2013.

[7] Hayat. S, Javaid. N, Khan. Z. A, Shareef. A, Mahmood. A, Bouk. S. H, "Energy Efficient MAC Protocols in Wireless Body Area Sensor Networks", 5th International Symposium on Advances of High Performance Computing and Networking (AHPCN-2012) in conjunction with 14th IEEE International Conference on High Performance Computing and Communications (HPCC-2012), 25-27 June, Liverpool, UK, 2012.

[8] Rahim. A, Javaid. N, Aslam. M, Qasim. U, Khan. Z. A, "Adaptive-Reliable Medium Access Control Protocol for Wireless Body Area Networks", Poster Session of 9th Annual IEEE Communications Society Conference on Sensor, Mesh and Ad Hoc Communications and Networks (SECON2012), Seoul, Korea, 2012.

[9] A. N Alvi, S. S. Naqvi, S. H. Bouk, N. Javaid, U. Qasim, Z. A. Khan, "Evaluation of Slotted CSMA/CA of IEEE 802.15.4", Broadband, Wireless Computing, Communication and Applications (BWCCA), 2012 Seventh International Conference on, vol., no., pp.391,396, 12-14 Nov. 2012.

[10] A. Biroli, M. Martina, and G. Masera, "An ldpc decoder architecture for wireless sensor network applications", Sensors, vol. 12, no. 2, pp. 15291543, 2012.

[11] P. Grover, K. Woyach, and A. Sahai, "Towards a communication-theoretic understanding of system-level power consumption", Selected Areas in Communications, IEEE Journal on, vol. 29, no. 8, pp. 17441755, 2011.

[12] L. Howard Sheryl, S. Christian, I. Kris, and I. Kris, "Error control coding in low-power wireless sensor networks: When is ecc energy-efficient", EURASIP Journal on Wireless Communications and Networking, vol. 2006.

[13] M. Pellenz, R. Souza, and M. Fonseca, "Error control coding in wireless sensor networks", Telecommunication Systems, vol. 44, no. 1, pp. 6168, 2010.

[14] Z. Kashani and M. Shiva, "Power optimised channel coding in wireless sensor networks using low-density parity-check codes", Communications, IET, vol. 1, no. 6, pp. 12561262, 2007.

[15] Y. Hamada, K. Takizawa, and T. Ikegami, "Highly reliable wireless body area network using error correcting codes", in Radio and Wireless Symposium (RWS), 2012 IEEE, pp. 231234, IEEE, 2012.

[16] J. Abouei, J. Brown, K. Plataniotis, and S. Pasupathy, "Energy eciency and reliability in wireless biomedical implant systems", Information Technology in Biomedicine, IEEE Transactions on, vol. 15, no. 3, pp. 456466, 2011.

[17] Z. Cai, J. Hao, and L.Wang, "An efficient early stopping scheme for ldpc decoding based on check-node messages", in Communication Systems, 2008. ICCS 2008. 11th IEEE Singapore International Conference on, pp. 13251329, IEEE, 2008.

[18] Javaid, N.; Mohammad, S.N.; Latif, K.; Qasim, U.; Khan, Z.A.; Khan, M.A., "HEER: Hybrid Energy Efficient Reactive protocol for Wireless Sensor Networks," Electronics, Communications and Photonics Conference (SIECPC), 2013 Saudi International , vol., no., pp.1,4, 27-30 April 2013.

[19] Javaid, N.; Khan, Z.A.; Qasim, U.; Khan, M.A.; Latif, K.; Javaid, A., "Towards LP modeling for maximizing throughput and minimizing routing delay in proactive protocols in Wireless Multi-hop Networks," Electronics, Communications and Photonics Conference (SIECPC), 2013 Saudi International , vol., no., pp.1,4, 27-30 April 2013.

[20] T M. Tahir, N. Javaid, A. Iqbal, Z. A. Khan, N. Alrajeh, "On Adaptive Energy Efficient Transmission in WSNs", International Journal of Distributed Sensor Networks, Volume 2013 (2013), Article ID 923714, 10 pages http://dx.doi.org/10.1155/2013/923714.

[21] Javaid, N.; Khan, A.A.; Akbar, M.; Khan, Z.A.; Qasim, U., "SRP-MS: A new routing protocol for delay tolerant Wireless Sensor Networks," Electrical and Computer Engineering (CCECE), 2013. 26th Annual IEEE Canadian Conference on , vol., no., pp.1,4, 5-8 May 2013.

[22] Ahmad, A.; Latif, K.; Javaidl, N.; Khan, Z.A.; Qasim, U., "Density controlled divide-and-rule scheme for energy efficient routing in Wireless Sensor Networks," Electrical and Computer Engineering (CCECE), 2013 26th Annual IEEE Canadian Conference on , vol., no., pp.1,4, 5-8 May 2013.

[24] Rasheed, M.B.; Javaid, N.; Khan, Z.A.; Qasim, U.; Ishfaq, M., "E-HORM: An energy-efficient

hole removing mechanism in Wireless Sensor Networks," Electrical and Computer Engineering (CCECE), 2013 26th Annual IEEE Canadian Conference on , vol., no., pp.1,4, 5-8 May 2013.

[25] Kashaf, A.; Javaid, N.; Khan, Z.A.; Khan, I.A., "TSEP: Threshold-Sensitive Stable Election Protocol for WSNs," Frontiers of Information Technology (FIT), 2012 10th International Conference on , vol., no., pp.164,168, 17-19 Dec. 2012.

[26] Ahmed, S.H.; Bouk, S.H.; Javaid, N.; Sasase, I., "Combined Human, Antenna Orientation in Elevation Direction and Ground Effect on RSSI in Wireless Sensor Networks," Frontiers of Information Technology (FIT), 2012 10th International Conference on , vol., no., pp.46,49, 17-19 Dec. 2012.

[27] Abbas, Z.; Javaid, N.; Khan, M.A.; Ahmed, S.; Qasim, U.; Khan, Z.A., "Simulation Analysis of IEEE 802.15.4 Non-beacon Mode at Varying Data Rates," Broadband, Wireless Computing, Communication and Applications (BWCCA), 2012 Seventh International Conference on , vol., no., pp.46,52, 12-14 Nov. 2012.

[28] Ain, Q.; Ikram, A.; Javaid, N.; Qasim, U.; Khan, Z.A., "Modeling Propagation Characteristics for Arm-Motion in Wireless Body Area Sensor Networks," Broadband, Wireless Computing, Communication and Applications (BWCCA), 2012 Seventh International Conference on , vol., no., pp.186,191, 12-14 Nov. 2012

[29] Kumar, S.; Javaid, N.; Yousuf, Z.; Kumar, H.; Khan, Z.A.; Qasim, U., "On link availability probability of routing protocols for urban scenario in VANETs," Open Systems (ICOS), 2012 IEEE Conference on , vol., no., pp.1,6, 21-24 Oct. 2012.

[30] Sagar, S.; Javaid, N.; Saqib, J.; Khan, Z.A.; Qasim, U.; Khan, M.A., "On probability of link availability in original and modified AODV, FSR and OLSR using 802.11 and 802.11p," Open Systems (ICOS), 2012 IEEE Conference on , vol., no., pp.1,6, 21-24 Oct. 2012.

[31] Qureshi, T.N.; Javaid, N.; Malik, M.; Qasim, U.; Khan, Z.A., "On Performance Evaluation of Variants of DEEC in WSNs," Broadband, Wireless Computing, Communication and Applications (BWCCA), 2012 Seventh International Conference on , vol., no., pp.162,169, 12-14 Nov. 2012.

[32] Anas, M.; Javaid, N.; Mahmood, A.; Raza, S. M.; Qasim, U.; Khan, Z. A., "Minimizing Electricity Theft Using Smart Meters in AMI," P2P, Parallel, Grid, Cloud and Internet Computing (3PGCIC), 2012 Seventh International Conference on , vol., no., pp.176,182, 12-14 Nov. 2012.

[33] Mahmood, D.; Javaid, N.; Qasim, U.; Khan, Z.A., "Routing Load of Route Calculation and Route Maintenance in Wireless Proactive Routing Protocols," Broadband, Wireless Computing, Communication and Applications (BWCCA), 2012 Seventh International Conference on , vol., no., pp.149,155, 12-14 Nov. 2012.

[34] Manzoor, B.; Javaid, N.; Bibi, A.; Khan, Z. A.; Tahir, M., "Noise Filtering, Channel Modeling and Energy Utilization in Wireless Body Area Networks," High Performance Computing and Communication & 2012 IEEE 9th International Conference on Embedded Software and Systems (HPCC-ICESS), 2012 IEEE 14th International Conference on , vol., no., pp.1754,1761, 25-27 June 2012.

[35] Khan, N. A.; Javaid, N.; Khan, Z. A.; Jaffar, M.; Rafiq, U.; Bibi, A., "Ubiquitous HealthCare in Wireless Body Area Networks," Trust, Security and Privacy in Computing and Communications (TrustCom), 2012 IEEE 11th International Conference on , vol., no., pp.1960,1967, 25-27 June 2012.

[36] ur Rehman, O.; Javaid, N.; Bibi, A.; Khan, Z.A., "Performance Study of Localization Techniques

in Wireless Body Area Sensor Networks," Trust, Security and Privacy in Computing and Communications (TrustCom), 2012 IEEE 11th International Conference on , vol., no., pp.1968,1975, 25-27 June 2012.

[37] Hayat, S.; Javaid, N.; Khan, Z. A.; Shareef, A.; Mahmood, A.; Bouk, S.H., "Energy Efficient MAC Protocols," High Performance Computing and Communication & 2012 IEEE 9th International Conference on Embedded Software and Systems (HPCC-ICESS), 2012 IEEE 14th International Conference on , vol., no., pp.1185,1192, 25-27 June 2012.

[38] Javaid, N.; Bibi, A.; Javaid, A.; Malik, S.A., "Modeling routing overhead generated by wireless proactive routing protocols," GLOBECOM Workshops (GC Wkshps), 2011 IEEE , vol., no., pp.1072,1076, 5-9 Dec. 2011.

[39] Javaid, N.; Bibi, A.; Djouani, K., "Interference and bandwidth adjusted ETX in wireless multi-hop networks," GLOBECOM Workshops (GC Wkshps), 2010 IEEE , vol., no., pp.1638,1643, 6-10 Dec. 2010.

[40] Javaid, N.; Javaid, A.; Khan, I.A.; Djouani, K., "Performance study of ETX based wireless routing metrics," Computer, Control and Communication, 2009. IC4 2009. 2nd International Conference on , vol., no., pp.1,7, 17-18 Feb. 2009